\documentclass[twocolumn]{aastex63}
\usepackage{graphicx}
\usepackage{color}
\usepackage{amsmath,amssymb}
\usepackage{gensymb}
\usepackage{times}
\usepackage{booktabs}
\usepackage{nicefrac}

\usepackage[utf8]{inputenc}

\begin{document}
\title{Chemical Cartography of the Sagittarius Stream with Gaia}
\author[0000-0002-6993-0826]{Emily C. Cunningham}
\altaffiliation{NASA Hubble Fellow}
\affiliation{Department of Astronomy, Columbia University, 550 West 120th Street, New York, NY, 10027, USA}
\affiliation{Center for Computational Astrophysics, Flatiron Institute, 162 5th Ave, New York, NY 10010, USA}

\author[0000-0001-8917-1532]{Jason A.S. Hunt}
\affiliation{Center for Computational Astrophysics, Flatiron Institute, 162 5th Ave, New York, NY 10010, USA}

\author[0000-0003-0872-7098]{Adrian M. Price-Whelan}
\affiliation{Center for Computational Astrophysics, Flatiron Institute, 162 5th Ave, New York, NY 10010, USA}

\author[0000-0001-6244-6727]{Kathryn V. Johnston}
\affiliation{Department of Astronomy, Columbia University, 550 West 120th Street, New York, NY, 10027, USA}

\author[0000-0001-5082-6693]{Melissa K. Ness}
\affiliation{Department of Astronomy, Columbia University, 550 West 120th Street, New York, NY, 10027, USA}
\affiliation{Center for Computational Astrophysics, Flatiron Institute, 162 5th Ave, New York, NY 10010, USA}

\author[0000-0003-4769-3273]{Yuxi (Lucy) Lu}
\affiliation{Department of Astronomy, Columbia University, 550 West 120th Street, New York, NY, 10027, USA}
\affiliation{Department of Astrophysics, American Museum of Natural History, 200 Central Park West, Manhattan, NY, USA}

\author[0000-0002-9933-9551]{Ivanna Escala}
\altaffiliation{Carnegie-Princeton Fellow}
\affiliation{Department of Astrophysical Sciences, Princeton University, 4 Ivy Lane, Princeton, NJ 08544, USA}
\affiliation{The Observatories of the Carnegie Institution for Science, 813 Santa Barbara St, Pasadena, CA 91101, USA}

\author[0000-0002-4110-8769]{Ioana A. Stelea}
\affiliation{Department of Astronomy, Columbia University, 550 West 120th Street, New York, NY, 10027, USA}

\correspondingauthor{Emily C. Cunningham}
\email{e.cunningham@columbia.edu}

\begin{abstract}

The stellar stream connected to the Sagittarius (Sgr) dwarf galaxy is the most massive tidal stream that has been mapped in the Galaxy, and is the dominant contributor to the outer stellar halo of the Milky Way. We present metallicity maps of the Sgr stream, using 34,240 red giant branch stars with inferred metallicities from Gaia BP/RP spectra. This sample is larger than previous samples of Sgr stream members with chemical abundances by an order of magnitude. We measure metallicity gradients with respect to Sgr stream coordinates $(\Lambda, B)$, and highlight the gradient in metallicity with respect to stream latitude coordinate $B$, which has not been observed before. We find $\nabla \mathrm{[M/H]} = -2.48 \pm 0.08 \times 10^{-2}$ dex/deg above the stream track ($B>B_0$ where $B_0=1.5 \degree$ is the latitude of the Sgr remnant) and $\nabla \mathrm{[M/H]} =- 2.02 \pm 0.08 \times 10^{-2}$ dex/deg below the stream track ($B<B_0$). By painting metallicity gradients onto a tailored N-body simulation of the Sgr stream, we find that the observed metallicities in the stream are consistent with an initial radial metallicity gradient in the Sgr dwarf galaxy of $\sim -0.1$ to $-0.2$ dex/kpc, well within the range of observed metallicity gradients in Local Group dwarf galaxies. Our results provide novel observational constraints for the internal structure of the dwarf galaxy progenitor of the Sgr stream. Leveraging new large datasets in conjunction with tailored simulations, we can connect the present day properties of disrupted dwarfs in the Milky Way to their initial conditions.
    
\end{abstract}


\section{Introduction}

Wrapped around the Milky Way (MW) are the leading and trailing arms of the Sagittarius (Sgr) stream, the most massive semi-preserved relic of the MW's hierarchical assembly. The leading and trailing arms connect at the remnant of their dwarf galaxy progenitor, the Sgr dwarf galaxy (\citealt{Ibata1994}). Debris from this system is scattered all over the sky and over Galactocentric distances ranging from $\sim 15$ kpc to $\sim 130$ kpc (e.g., \citealt{Ibata2001} \citealt{Majewski2003}, \citealt{Belokurov2014},  \citealt{Hernitschek2017}, \citealt{Law2016}, \citealt{Sesar2017}). The unraveling of the Sgr dwarf galaxy into the Sgr stream is believed to have begun approximately 6 Gyr ago (e.g., \citealt{Laporte2018}), when the Sgr dwarf galaxy progenitor is thought to have first crossed the MW virial radius and began undergoing tidal disruption by the MW. The Sgr system presents the unique opportunity to explore what debris from ancient accretion events might tell us about their dwarf galaxy progenitors, in a case where the progenitor remnant is very much alive and observable.

Because its debris extends over such a large radial range within the Galaxy, the Sgr stream can be used to constrain the potential of the MW, including the mass and shape of its dark matter halo (e.g., \citealt{Helmi2004}, \citealt{Law2005}, \citealt{Johnston2005}, \citealt{LM2010}, \citealt{Deg2013}, \citealt{VeraCiro2013}, \citealt{Ibata2013}, \citealt{Gibbons2014}, \citealt{Law2016} \citealt{Dierickx2017}, \citealt{Fardal2019}, \citealt{Vasiliev2021}, \citealt{Panithanpaisal2022}). An essential component of this work has been simulating the tidal disruption of Sgr to compare with the observations. However, estimates of the initial mass of the Sgr dwarf galaxy progenitor remain uncertain over an order of magnitude. The dwarf galaxy progenitor of the Sgr stream is thought to have been a relatively massive, luminous satellite ($L\sim 10^8 L_{\odot}$, $M_{\star} \sim 10^{8} M_{\odot}$ $M_{\rm tot} > 10^9 M_{\odot}$; e.g., \citealt{Penarrubia2010}, \citealt{Niederste-Ostholt2012}, \citealt{Deason2019}), though several studies have argued for an even more massive Sgr progenitor, comparable to the Large Magellanic Cloud  ($M_{\rm tot}>6 \times 10^{10}- 1 \times 10^{11} M_{\odot}$ e.g., \citealt{Gibbons2017}, \citealt{Laporte2018}, \citealt{Mucciarelli2017}).  Constraining the properties of the Sgr dwarf galaxy progenitor is critical both for using the debris to constrain the MW's mass distribution and for understanding Sgr's role in contributing to disequilibrium in the MW. 

The chemical properties of stars in the Sgr remnant and stream provide further information about the Sgr dwarf galaxy progenitor. Stellar mass in dwarf galaxies correlates with mean metallicity (e.g., \citealt{Kirby2013}, \citeyear{Kirby2020}); estimating the mean metallicity of Sgr could therefore help constrain the mass of its dwarf galaxy progenitor. One complication is mounting evidence that the Sgr dwarf galaxy progenitor had a radial metallicity gradient prior to disruption, first suggested when stream stars were found to be, on average, lower metallicity than stars in the core of the Sgr dwarf spheroidal (e.g., \citealt{Chou2007}, \citealt{Monaco2007}). As samples grew, chemical differences along the different arms of the stream were observed (e.g., \citealt{Keller2010}, \citealt{Carlin2012}, \citealt{Shi2012}, \citealt{Hyde2015}, \citealt{Carlin2018}, \citealt{Hasselquist2019}, \citealt{Yang2019}, \citealt{Hayes2020}, \citealt{Zhao2020}, \citealt{Ramos2022}). Several studies have argued that there is a metallicity gradient in the core (e.g., \citealt{Majewski2013}, \citealt{Mucciarelli2017}, \citealt{Vitali2022}, \citealt{Minelli2023}). Other works have argued in favor of an initial metallicity gradient based on the radial velocity dispersions of stream stars in different metallicity bins (e.g., \citealt{Gibbons2017}, \citealt{Johnson2020}, \citealt{Limberg2023}). In Sgr, metal-poor stars have been observed to have a higher radial velocity dispersion than the metal-rich stars (though see \citealt{Penarrubia2021} for how sample selection can contribute in part to this finding). Using stars from the SEGUE survey, \cite{Limberg2023} showed that this enhanced velocity dispersion at lower metallicity is likely because the lower-metallicity populations have stars from multiple wraps of the stream (i.e., are at different orbital phases). 

 The Sgr system represents a special opportunity to study star formation in a dwarf galaxy on a star-by-star basis, and can be used to shed some light on the physical processes that are responsible for observed radial metallicity gradients in dwarf galaxies. Radial stellar metallicity gradients (of varying magnitudes) appear to be ubiquitous in Local Group dwarf galaxies, including satellites of the MW (e.g., \citealt{Harbeck2001}, \citealt{Battaglia2006}, \citeyear{Battaglia2011}, \citealt{Gullieuszik2009}, \citealt{Kirby2011}, \citealt{Dobbie2014}, \citealt{Spencer2017}, \citealt{Pace2020}), M31 (\citealt{Vargas2014}, \citealt{Ho2015}) and isolated dwarfs (\citealt{Leaman2013}, \citealt{Kirby2017}, \citealt{Taibi2018}, \citeyear{Taibi2020}, \citealt{Hermosa2020}, \citealt{Choudhury2016}, \citeyear{Choudhury2018}, \citeyear{Choudhury2021}). In simulations, metallicity gradients in dwarf galaxies can arise due to preferential star formation in the central regions (e.g., \citealt{Schroyen2013}), as well as stellar feedback and dynamical heating gradually resulting in older populations moving outwards in time (creating a correlation with age; e.g., \citealt{Mercado2021}). However, observationally, which galaxy properties correlate with the slope of this gradient remains unclear. \cite{Taibi2022} recently compiled a comprehensive sample of red giant branch stars in dwarf galaxies with measured chemical abundances, and found no correlation with metallicity gradient slope with age, dynamical state or star formation history. External perturbations can also have effects: the Sgr system is particularly interesting from this perspective, as it has been undergoing tidal disruption by the MW and star formation concurrently over the last $\sim 6$ Gyr, with stellar populations as young as $\sim 1-3$ Gyr found in the central nucleus (\citealt{Siegel2007}, \citealt{Alfaro2019}). 

Plaguing all of the existing chemical studies of the Sgr stream with spectroscopic metallicites are: 1) incomplete coverage along and across the stream and 2) relatively small numbers of stars (with early studies having $\sim$10 stars; more recently, typical sample sizes are a few hundred stars in each arm). In addition, metallicity biases in sample selections can also be an issue in inferring the initial metallicity gradient of the progenitor (e.g., see \citealt{Limberg2023}'s discussion of the metal-poor selection bias in SEGUE). Finally, most studies of metallicity in the Sgr stream have focused on trends with stream longitude (along the stream), and not with stream latitude (across the stream). No gradient has been detected across the stream, though a gradient perpendicular to the stream track has been detected in Andromeda's Giant Stellar Stream (\citealt{Ibata2007}, \citealt{Escala2021}).

To fully map the stream with chemical abundances, we require a large, all-sky sample of Sgr stream candidates as free from metallicity biases as possible. This is now possible for the first time thanks to the third data release from the Gaia mission \citep[Gaia DR3;][]{GaiaMission2016, GaiaDR3}. One key data product from Gaia DR3 are the 220 million low-resolution ($R \sim 40-150$) spectra, taken with the BP and RP prisms (\citealt{Carrasco2021}, \citealt{DeAngeli2022}), which we hereafter refer to as Gaia XP spectra. While an initial catalog of stellar parameter estimates derived from synthetic spectra was released along with Gaia DR3 \citep{Andrae2022}, this catalog has been shown to suffer from systematics and catastrophic failures, in part due to our presently imperfect understanding of the XP system as well as the fact that low-resolution spectra will not be equally informative for all stellar types. However, \cite{Rix2022} and \cite{Andrae2023} demonstrate that data-driven metallicity estimates derived from the XP spectra have more success when validated against other surveys, particularly for cooler stars (see also the catalogs from \citealt{Zhang2023}, \citealt{Yao2023}). Given the tremendous number of stars with stellar parameter estimates from Gaia XP, as well as the improvement of the data-driven estimates for describing metal-poor stars, this dataset is ideal for mapping the chemical structure of the Sgr stream. 

In this paper, we investigate the chemical properties of the Sgr stream using $\sim$ 34,000 red giant branch (RGB) stars (selected in \citealt{Vasiliev2021}) with data-driven inferred metallicities from the Gaia XP spectra (measured in \citealt{Andrae2023}). In Section \ref{sec:data}, we introduce the datasets used in this work. In Section \ref{sec:methods}, we discuss our Bayesian mixture modeling approach to estimate metallicity gradients in the stream, and summarize the our results from our inference procedure in Section \ref{sec:results}.
We interpret our findings in the context of an N-body simulation of a Sgr-like stream in Section \ref{sec:sims}, and summarize our findings in Section \ref{sec:concl}. 

\begin{figure*}
\centering
    \includegraphics[width=\textwidth]{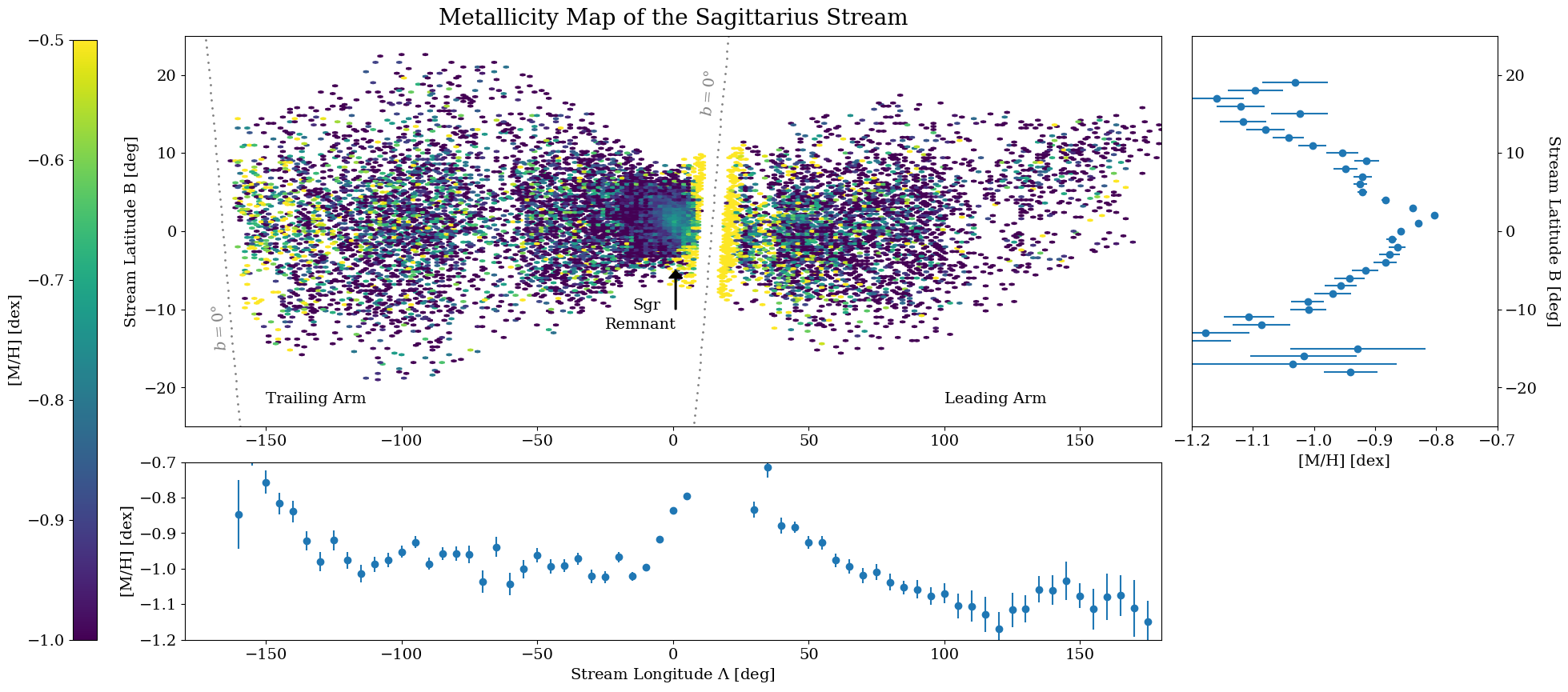}
    \caption{Metallicity map of the Sgr stream. Main panel: positions of high probability Sgr stream members (selected in V21) are shown in Sgr stream coordinates $(\Lambda, B)$, color-coded by the mean [M/H] value as measured in A23. MW disk contaminants in the V21 sample are visible as high metallicity stars near the Galactic midplane (shown by the dotted gray line). The total sample consists of $\sim 34$k stars with metallicity estimates, greater than previous samples by an order of magnitude. Lower panel: mean metallicity binned in stream longitude $\Lambda$. Errorbars are the uncertainty on the mean in each bin. We note that the metallicity range shown here excludes the high metallicity stars near the Galactic midplane, in order to emphasize the metallicity trends along the stream; we refer the interested reader to Figure \ref{fig:1dres} for the full metallicity range spanned by the data. Right panel: mean metallicity binned in stream latitude $B$. There is a clear gradient in metallicity with respect to $|B-B_0|$, which has not been observed before. }
    \label{fig:map}    
\end{figure*}

\begin{figure}
    \centering
    \includegraphics[width=0.49\textwidth]{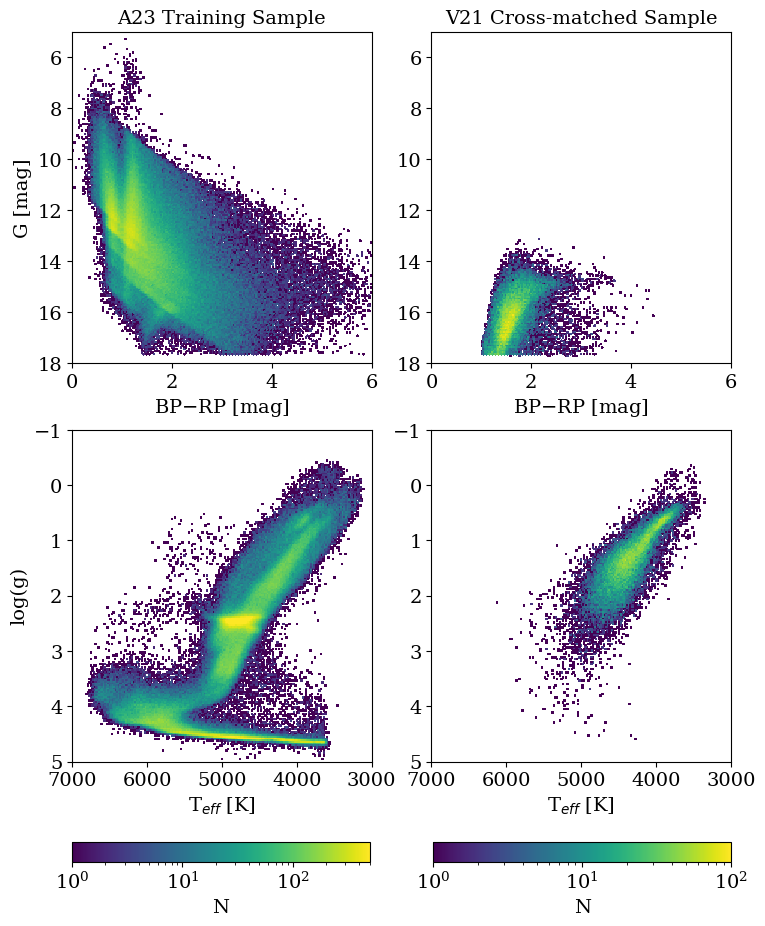}
    \caption{Top panels: color-magnitude diagrams showing the range of colors and apparent magnitudes of stars in the A23 training sample (left) and the V21--A23 cross-matched sample (right). Lower panels: Hertzsprung-Russell diagrams of the A23 full training sample (left) and the V21--A23 cross-matched sample (right). While the stars studied in this work are at the fainter end of stars with published estimates based on the Gaia XP spectra, they are also relatively cool red giants, and within the regime for which the derived metallicities from Gaia XP spectra are expected to be reliable (see Section \ref{subsec:a23} and A23 for more detail).}
    \label{fig:hr}
\end{figure}

\begin{figure}
    \centering
    \includegraphics[width=0.4\textwidth]{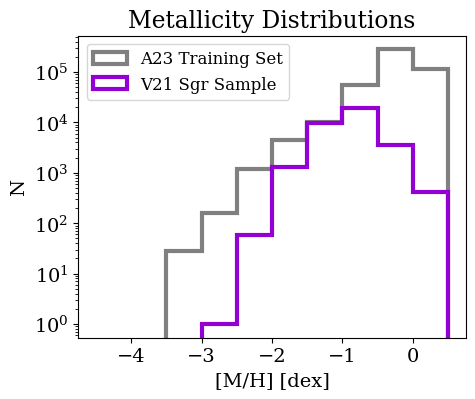}
    \caption{Metallicity distribution function for the A23 training set (gray) and the Sgr sample presented here (purple), shown on a logarithmic scale. The metallicities of stars similar to those in the Sgr sample are well represented by the A23 training set. }
    \label{fig:training_mdfs}
\end{figure}

\begin{figure}
    \centering
    \includegraphics[width=0.4\textwidth]{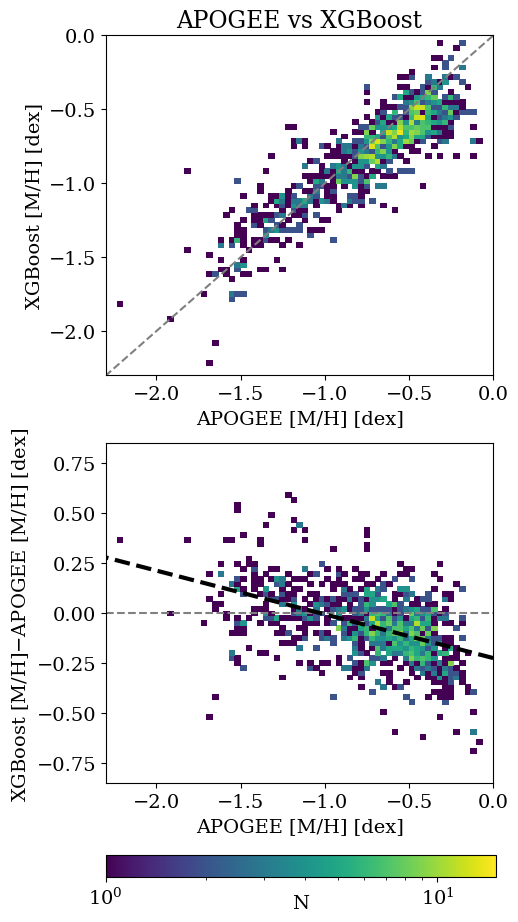}
    \caption{Estimates of [M/H] for stars in the V21 sample with measurements from both A23 and APOGEE DR17. We generally find good agreement between the two samples of [M/H] estimates. We note a slight bias in the A23 estimates relative to the APOGEE measurements, where the XGBoost metallicities are slightly underestimated at high metallicity and overestimated at lower metallicity compared to the APOGEE reported values (black dashed line in the right panel). However, the consequence of this difference is that the true gradients in the Sgr stream may be stronger than those reported here.}
    \label{fig:apogee_v_XGBoost}
\end{figure}

\section{Data}
\label{sec:data}

In Figure \ref{fig:map}, we present a metallicity ([M/H]) map of the Sgr stream, constructed from $\sim$ 34,000 red giant branch stars that are likely Sgr members (selected in \citealt{Vasiliev2021}) with metallicities inferred from Gaia XP spectra (measured in \citealt{Andrae2023}). In this section, we introduce the datasets used to construct this sample. We first define the coordinate system that we use for the remainder of this paper in Section \ref{subsec:coords}; we discuss the catalog of likely Sgr members in Section \ref{subsec:v21} and their Gaia XP inferred metallicity estimates in Section \ref{subsec:a23}.

\subsection{Sgr Stream Coordinate System}
\label{subsec:coords}

For the purposes of this work, we use the right-handed coordinate system ($\Lambda, B$) for the Sgr stream, as defined in \cite{Vasiliev2021}. In this coordinate system, the Sgr remnant is located at $(\Lambda_0, B_0)\approx (0 \deg,1.5 \deg)$. This coordinate system uses the $\Lambda$ convention introduced in \cite{Belokurov2014}, in which the leading arm has $\Lambda>0$ and the trailing arm has $\Lambda<0$, but with $B$ defined as in \cite{Majewski2003}, such that the coordinate system is right-handed. 

\subsection{Candidate Sgr Stream Members}
\label{subsec:v21}

In this paper, we make use of a publicly available sample of red giant branch (RGB) stars that are high probability Sgr stream members, published in \cite{Vasiliev2021} (hereafter V21).\footnote{V21 catalog available at \url{https://zenodo.org/record/4038137}} This sample was constructed by first cross-matching 2MASS \citep{Skrutskie2006AJ} with Gaia DR2 \citep{GaiaDR2}, keeping only stars with Galactic latitudes $|b|>5$, extinction-corrected $13 < G < 18$, $BP- RP > 1$ and parallax $\varpi < 0.1$ mas. Likely Sgr stream members were selected from this cross-matched catalog based off their proper motions, colors (in both \textit{Gaia} and 2MASS bands), and absolute magnitudes. V21 estimate absolute magnitudes (and therefore distances) to each star in their sample by using RR Lyrae from \textit{Gaia} DR2 to calculate the distance modulus of the stream as a function of stream longitude $\Lambda$. We refer the interested reader to Section 2.3 and Figure 1 of V21 for additional details on their selection process. While there are other available catalogs that kinematically select Sgr stream members (e.g., \citealt{Antoja2020}, \citealt{Ramos2022}), we use the V21 sample here in particular because they select RGB stars, for which we expect the Gaia XP abundances to be the most reliable (see, e.g., Section 4.3 and Figure 9 of A23). 

\subsection{Abundances from Gaia XP Spectra}
\label{subsec:a23}

We use stellar element abundance estimates inferred from the Gaia XP spectra as reported in the publicly-available catalog from \cite{Andrae2023}\footnote{A23 catalog available at: \url{https://zenodo.org/record/7599789}. At the time of writing this publication, the latest available version of the catalog is Version 2.1.} (hereafter A23). 
These metallicity ([M/H]) values are derived from training XGBoost models \citep{Chen2016} to estimate [M/H], surface gravity, and effective temperature from the Gaia XP spectra as well as synthesized photometry computed from GaiaXPy \citep{GaiaPhot2023}. We refer the reader to Appendix A of A23 for the full list of inputs and implementation for these XGBoost models. 
The models are trained on stars from APOGEE DR17 \citep{Abdurro'uf2022} augmented by a sample of very metal poor stars from \cite{Li2022}, in order to cover the full metallicity range of stars in the Local Group. In order to qualify for the training sample, stars must have high resolution spectra with S/N$>$50, available Gaia XP spectra, and CatWISE photometry \citep{Marocco2020}.

The color-magnitude diagram for the full training sample, along with the inferred surface gravities and effective temperatures, are shown in the left panels of Figure \ref{fig:hr}; the stars from the V21/A23 cross-match are shown in the right panels. The metallicity distribution functions (MDFs) for both the training and V21/A23 cross-match are shown in Figure 3. While the stars in the Sgr sample are at the faint end of the Gaia XP sample, the color and metallicity ranges of the V21/A23 cross-match are well represented in the training set, and A23 find that their estimates are robust (particularly for cooler RGB stars, as we have here) down to the faint limit of the Gaia XP sample (see their Figures 4d and 8). We refer the reader to A23 for details on the extensive validation of this dataset.   

While uncertainties of metallicity estimates for individual stars are not reported in A23, based on their model validation, they report an accuracy of $0.1$ dex. Therefore, in our analysis below, we assume each [M/H] estimate has a measurement uncertainty of $\sigma_{\rm err} =0.1$ dex. In Section \ref{sec:methods}, we leave the metallicity dispersion of Sgr ($\sigma_{\rm Sgr}$) as a free parameter, and note that this dispersion includes both the intrinsic scatter and any additional measurement uncertainty beyond the assumed 0.1 dex.

In order to further validate our use of this dataset, we compare the A23 metallicity measurements for the 1394 stars in V21 that have abundances reported in APOGEE DR17 \citep{Abdurro'uf2022} in Figure \ref{fig:apogee_v_XGBoost}. As seen in the left panel of Figure \ref{fig:apogee_v_XGBoost}, we find generally good agreement between the two samples, consistent with the validation in A23. In the right panel of Figure \ref{fig:apogee_v_XGBoost}, we compare the differences between the APOGEE and A23 abundances as a function of their APOGEE metallicities. We do see that the XGBoost estimates from A23 are slightly too low compared to the APOGEE estimates at higher metallicity, while they are slightly too high at low metallicity. This is also noted in A23 (see Figure 4 and Section 2.4 of A23). A23 suggests that this is could be due to the fact that they used the [Fe/H] values measured in \citep{Li2022} for the metal-poor stars in their training sample, while they used the [M/H] values reported by APOGEE. While a deeper investigation of this result is beyond the scope of this work, we note that these differences only serve to strengthen the underlying gradients detected in Section \ref{sec:results}. Because the XGBoost metallicites may be underestimated at higher metallicities (e.g., in the Sgr core) and overestimated at low metallicities (e.g., at the edges of the stream), the true underlying metallicity gradients may be stronger than those reported here. Larger samples of higher resolution spectroscopic metallicities of stars in Sgr (e.g., from large surveys such as DESI; \citealt{DESI:2016}) are needed to determine if our gradients are underestimated.

The resulting cross-matched sample shown Figures \ref{fig:map} and \ref{fig:hr} consists of 34,240 stars. This sample is larger than previous samples of stars with chemical abundance estimates in the Sgr stream by an order of magnitude (though we note that there are large samples of Sgr main body stars that have estimates from photometric metallicities; e.g., \citealt{Vitali2022}). The Sgr main body can be seen in Figure \ref{fig:map} at $(\Lambda, B)=(\Lambda_0, B_0)\sim(0\deg,1.5\deg)$; there is a local peak in the metallicity map at the core of $\sim -0.8$ dex and the mean metallicity visibly decreases in both coordinates as we move away from the center of the main body. The trends of metallicity with stream coordinates are shown in the lower and right panels of Figure \ref{fig:map}; the gradient with respect to stream latitude $B$, detected here for the first time, is immediately apparent in the binned data. We can also immediately see the contamination in the sample from metal-rich stars from the disk, which are located right along the edge of the Galactic latitude cut imposed by V21 (seen as the gap at $\Lambda \sim 10 \deg$). We make this cross-matched catalog, along with the membership probabilities computed in Section \ref{sec:results}, publicly available to the community.\footnote{Cross-matched catalog: \url{https://zenodo.org/record/8147032}.}

While we do not use the kinematical measurements in our analysis presented here, we include the updated Gaia DR3 \citep{GaiaDR3} astrometry and radial velocities (where available) with the publicly released catalog. While one could in principle re-do the V21 selection with Gaia DR3 astrometry, because the focus of this work are stars with Gaia XP spectra ($G<17.6$ mag), we use the original V21 selection, as the most significant improvements to the astrometry from Gaia DR2 to Gaia DR3 were at the fainter end of the sample. 
There are 7,685 stars in the cross-match with measured radial velocities. Of these, the vast majority follow a coherent, narrow track in radial velocity, except for the high metallicity stars near the midplane (i.e., the MW disk contaminants), which have a significantly higher radial velocity dispersion than the rest of the stars. We also note there are only a few lower-metallicity outliers along the stream in this sample, which could potentially members of previous wraps (see, e.g., \citealt{Limberg2023}) or field halo stars. We leave a deeper exploration of the full dataset with kinematics and chemistry to future work. 

In the next section, we describe our method for inferring metallicity gradients along and across the Sgr stream with this sample, as well as how we model the contamination from the MW's disk and other outliers. The results of our analysis are presented in Section \ref{sec:results}.

\section{Bayesian Mixture Modeling Procedure}
\label{sec:methods}

\begin{figure}
\centering
     \includegraphics[width=0.5\textwidth]{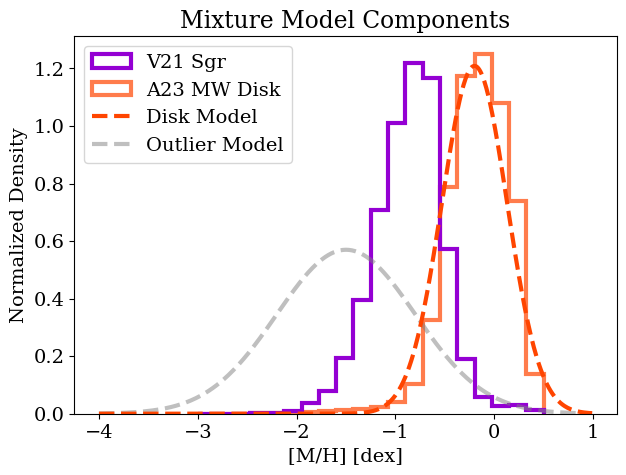}
     \caption{The components of our mixture model. The metallicity distribution for the V21 sample is shown in purple. Our disk model, derived from the A23 high fidelity RGB sample with $|b|<10 \degree$, is shown in orange, and our outlier model is shown in gray.}
     \label{fig:model}
\end{figure}

\begin{figure}
    \centering
    \includegraphics[width=0.45\textwidth]{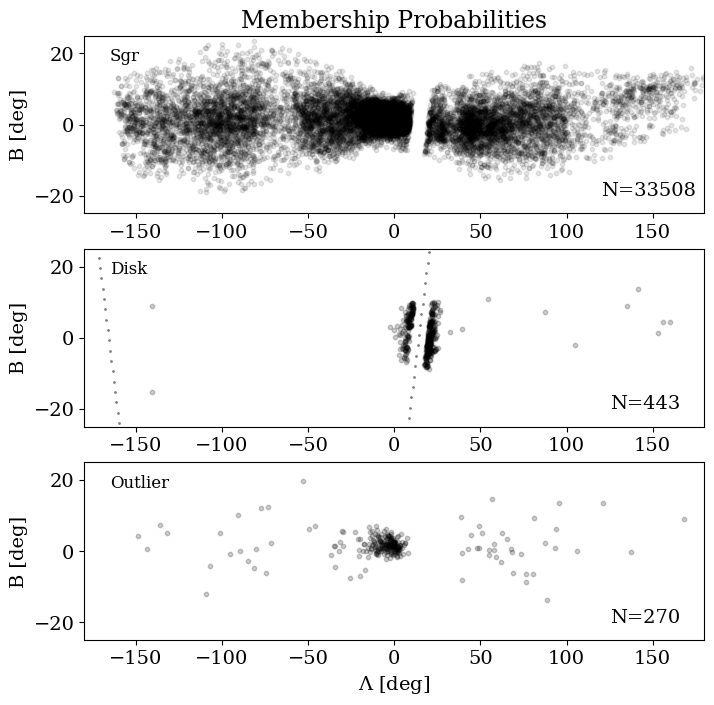}
    \caption{Maps of the V21 Sgr sample separated based on membership probability to the different components of our mixture model. The top panel shows all stars with Sgr membership probability greater than 0.5; this sample includes 33,508 stars. The middle panel shows stars with disk membership probabilities greater than 0.5; these are clustered near the Galactic midplane (indicated by the dotted line). The final panel shows stars with probabilities greater than 0.5 of belonging to the outlier distribution. These stars are distributed isotropically throughout the stream, but there is also a cluster near ($\Lambda_0,B_0$), as this component of the model is also picking up stars in the metal-poor tail of the remnant.}
    \label{fig:mem_prob}
\end{figure}

In this section, we describe our modeling procedure for estimating the metallicity gradients in the Sgr stream. From Figure \ref{fig:map}, it is immediately apparent that high metallicity stars from the disk have contaminated the V21 PM-selected catalog, as indicated by the metal-rich stars at low galactic latitudes. Furthermore, we want to account for the fact the catalogs may suffer from contamination from other halo stars that were not accreted from the Sgr dwarf galaxy. We therefore construct a Bayesian mixture model with three components: a component for the Sgr stream, a disk component, and an outlier component.

\textbf{Disk Model:} To determine the parameters of our disk model, we use the catalog of 17.5 million RGB stars with 3D kinematics and vetted parameter estimates from A23. The orange histogram in of Figure \ref{fig:model} shows the metallicity estimates of all high-fidelity RGB stars from this catalog within 10 degrees of the galactic midplane ($|b|<10 \degree$); we use the mean ($\mu_{\rm Disk}=-0.2$ dex) and standard deviation ($\sigma_{\rm [M/H], Disk}=0.33$ dex) of this sample as the parameters of the MDF of our disk population (orange curve in Figure \ref{fig:model}). 

We denote the full vector of observations $y=\{\mathrm{[M/H]}_1,...,\mathrm{[M/H]}_N \}$. The likelihood for a given observation $\mathrm{[M/H]}_i$ (where $i=1$ to $N$ and $N$ is the total number of stars) for the disk component of the model is given by:

\begin{equation}
p(\mathrm{[M/H]}_i|\mu_{\rm Disk}, \sigma_{\rm Disk})= \mathcal{N}(\mathrm{[M/H]}_i| \mu_{\rm Disk}, \sigma_{\rm Disk}^2+\sigma_{\rm err}^2), \\
\end{equation}
where $\mu_{\rm Disk}, \sigma_{\rm Disk}$ are fixed as seen in Figure \ref{fig:model}. 

\textbf{Outlier Model:} In addition to disk contaminants, we account for the fact that there many be halo stars in the V21 sample that were not accreted from the Sgr dwarf galaxy. In addition, metallicity distribution functions are known to not be Gaussian and have metal-poor tails; this outlier distribution also accounts for this. To account for non-Sgr halo interlopers and the metal-poor tail of the Sgr MDF, we incorporate a broad, metal-poor outlier component to our model. We assume that stars in the outlier component have a broad spread in metallicity (grey curve in right panels of Figure \ref{fig:model}; $\mu_{\rm Outlier}=-1.5$ dex, $\sigma_{\rm [Fe/H], Outlier}=0.7$ dex). This component of the model serves as a broad outlier distribution. 

The likelihood for a given observation ${\mathrm{[M/H]}_i}$ for the outlier component of the model is given by:

\begin{multline}
    p(\mathrm{[M/H]}_i|\mu_{\rm  Outlier}, \sigma_{\rm Outlier}) = \\
    \mathcal{N}(\mathrm{[M/H]}_i| \mu_{\rm  Outlier}, \sigma_{\rm  Outlier}^2+\sigma_{\rm err}^2),
\end{multline}
where, as with the disk component, $\mu_{\rm Outlier}, \sigma_{\rm Outlier}$ are fixed, and this component of the likelihood is not a function of model free parameters $\theta$. 

\textbf{Sgr Model}: For the Sgr component of our model, we consider three different parameterizations for the metallicity as a function of stream coordinates. In all cases, we model the metallicity of Sgr stars as being drawn from a normal distribution with mean $\mu_{\rm Sgr}$ and variance $\sigma_{\rm Sgr}^2$, but allow the mean $\mu_{\rm Sgr}$ to be a function of stream coordinates. 

First, we model the metallicity as a piecewise linear function of the stream longitude coordinate $\Lambda$, as done in previous work (e.g., \citealt{Hayes2020}). We allow the slopes of the trailing and leading arms (which we denote $\alpha_T$ and $\alpha_L$, respectively) to vary independently, but require that the two lines have the same value at the location of the progenitor $\Lambda_0=0$ (i.e., the function is continuous along the stream). In this model, the mean metallicity of the Sgr stream $\mu_{\rm Sgr}$ as a function of its stream coordinates is written as:  

\begin{equation}
\mu_{\rm Sgr}(\Lambda, B) = \mu_{\rm Sgr}(\Lambda) = \left\{
        \begin{array}{ll}
            \alpha_T \times |\Lambda| + C & \quad \Lambda \leq 0, \\
            \alpha_L \times \Lambda + C & \quad \Lambda > 0.
        \end{array}
    \right.
    \label{eqn:lambda_only}
\end{equation}
We fit the model in terms of $|\Lambda|$ for the trailing arm so that all of our gradient slopes are negative, reflecting the fact that the metallicity decreases as we move farther away from the Sgr core. The vector of free parameters that we estimate for the Sgr component of the model is $\theta_{\rm Sgr}= \{\alpha_T, \alpha_L, C, \sigma_{\rm Sgr}\}$.

We then model the metallicity as a piecewise linear function of the stream latitude coordinate $B$, with different slopes above and below the progenitor (located at $B_0=1.5 \degree$). We allow the slopes above the stream track and below the stream track to vary independently; we denote the slope above the stream track as $\beta_N$ and below the stream track as $\beta_S$. As above require that the two lines have the same value $C$ at $B_0$. For this model, the mean metallicity as a function of stream latitude is written as:

\begin{flalign}
\mu_{\rm Sgr}(\Lambda, B) & =\mu_{\rm Sgr}(B)  \\ \nonumber
& = \left\{
        \begin{array}{ll}
            \beta_N \times (B-B_0) + C & \quad B>B_0, \\
            \beta_S \times |B-B_0| + C & \quad B\leq B_0.
        \end{array}
    \right.
\end{flalign}
The free parameter vector for this Sgr model is thus $\theta_{\rm Sgr}= \{\beta_N, \beta_S, C, \sigma_{\rm Sgr}\}$.

Finally, we allow metallicity to vary as a linear function of both coordinates $(\Lambda, B)$. In this model, the mean [M/H] value $\mu_{\rm Sgr}$ is given by:

\begin{multline}
    \mu_{\rm Sgr}(\Lambda, B) = \\ \left\{
        \begin{array}{lc}
            \alpha_T |\Lambda| + \beta_N  (B-B_0) + C & \quad \Lambda \leq \Lambda_0, \\
            & \quad B>B_0,\\
            \alpha_T |\Lambda| + \beta_S  |B-B_0| + C & \quad \Lambda \leq \Lambda_0, \\
            & \quad B \leq B_0, \\
            \alpha_L \Lambda+ \beta_N (B-B_0)+ C & \quad \Lambda > \Lambda_0, \\
            & \quad B>B_0, \\
            \alpha_L \Lambda+ \beta_S |B-B_0|+ C & \quad \Lambda > \Lambda_0, \\
            & \quad B\leq B_0. 
        \end{array}
    \right.
    \label{eqn:lambda_beta}
\end{multline}
Our vector of free parameters for the Sgr component is therefore $\theta_{\rm Sgr}= \{\alpha_T, \alpha_L, \beta_N, \beta_S, C, \sigma_{\rm Sgr}\}$.

For every parameterization of the metallicity variation, we can write down the likelihood for the Sgr component of our model as:

\begin{multline}
    p(\mathrm{[M/H]}_i|\theta_{\rm Sgr}, \Lambda_i, B_i ) = \\
    \mathcal{N} (\mathrm{[M/H]}_i| \mu_{\rm Sgr}(\Lambda_i, B_i), \sigma_{\rm Sgr}^2+\sigma_{\rm err}^2),
\end{multline}
where $\mathcal{N}(x \,|\, \mu, \sigma)$ represents a normal distribution over $x$ with mean $\mu$ and standard deviation $\sigma$.

Our complete likelihood for a star's observed [M/H]$_i$ estimate, given stream coordinates $x_i=(\Lambda_i, B_i)$ and parameters $\theta=\{ f_{\rm Disk},f_{\rm Outlier}, \theta_{\rm Sgr}\}$ is given by:

\begin{multline}
    p(\mathrm{[M/H]}_i|\theta, \Lambda_i, B_i) = f_{\rm Disk} \times p(\mathrm{[M/H]}_i|\mu_{\rm Disk}, \sigma_{\rm Disk}^2)\\
    +f_{\rm Outlier} \times p(\mathrm{[M/H]}_i|\mu_{\rm Outlier}, \sigma_{\rm Outlier}^2) \\
    + (1-f_{\rm disk} -f_{\rm Outlier} ) \times p(\mathrm{[M/H]}_i|\theta_{\rm Sgr}, \Lambda_i, B_i).
\end{multline}

The posterior distribution is given by the product of the likelihoods for each star, multiplied by the prior distributions on our model parameters:

\begin{equation}
 p(\theta | y, x) = p(\theta) \prod_{i=0}^N p(\mathrm{[M/H]}_i | \theta, \Lambda_i, B_i) ,
\end{equation}
where we denote our full vector of metallicity measurements as $y=\{\mathrm{[M/H]}_1,...,\mathrm{[M/H]}_N \}$ and the full vector of coordinates as $x=\{(\Lambda_1, B_1), ..., (\Lambda_N, B_N) \}$.

We assume the non-informative \cite{Jeffreys1946} prior ($p(\sigma_{\rm Sgr}) \propto 1/\sigma_{\rm Sgr}$) on the dispersion of the Sgr component, and assume uniform priors on our fractions of disk and outlier contamination on the interval $[0,1]$ (and enforcing that their sum be less than one). We assume unbounded uniform priors on our gradient slopes (e.g., $p(\alpha_L) \propto $ const; $p(\beta_N) \propto $ const; $p(C) \propto $ const). We use \verb+emcee+ (\citealt{Foreman-Mackey2013}), the python implementation of the \cite{Goodman2010} Affine Invariant Markov Chain Monte Carlo (MCMC) Ensemble Sampler, to sample from our posterior distribution.

\section{Chemical Cartography}
\label{sec:results}

We present the main results from our analysis of the data in this section. Table \ref{tab:results} summarizes the results for each of our three models for the metallicity variation along and across the Sgr stream. We first demonstrate the effectiveness of our modeling approach by presenting membership probabilities in Section \ref{subsec:mem_prob}; we go on to show the trends (and measured gradients) with stream coordinates in Section \ref{subsec:trends}; and we compare the MDFs of the leading and trailing arms and discuss the Sgr bifurcation in Section \ref{subsec:bifurcation}. 

\subsection{Membership Probabilities}
\label{subsec:mem_prob}

While we discuss the results for our inferred gradients in the next subsection, we highlight first here the effectiveness of our mixture modeling approach. Figure \ref{fig:mem_prob} shows our sample split by membership probability to each component of our mixture model (using the results from our model where the mean metallicity is a function of both stream coordinates $\Lambda, B$). The top, middle and lower panels show stars with membership probability of greater than 0.5 to the Sgr, disk and outlier components of our model, respectively. Reassuringly, while the mixture model is only a function of the metallicities and does not contain any information about the star coordinates (except for the fact that the mean [M/H] is allowed to vary in the Sgr component of the model as a function of $(\Lambda, B)$), the populations with high disk and outlier probabilities are distributed as we would expect. The stars with the highest probability of being disk stars are clustered near the Galactic midplane (middle panel of Figure \ref{fig:mem_prob}, with the dotted line indicating the position of the midplane). While the stars with high probabilities of belonging to the outlier distribution are distributed isotropically throughout the stream (as one might expect from field halo contamination), there is also a cluster of metal-poor stars that are clearly members of the remnant. This demonstrates that a significant fraction of stars identified as belonging to the outlier component are stars in the metal poor tail of the Sgr MDF. A more sophisticated model allowing for the fact that the Sgr MDF is not Gaussian would likely improve this classification. For all models, the estimated disk contamination fraction is $\sim 2.5\%$ and the outlier contribution is $\sim 2.5\%$ (see Table \ref{tab:results}). We note that a more complex model that includes the spatial distributions of the different components, as well as incorporating kinematic information where available, would likely perform even better at removing non-Sgr contaminants. For the purposes of this work, we restrict ourselves to the simpler (yet still effective) approach. For reference, 33,508 stars (all but 732 stars) have Sgr membership probability greater than 0.5 in this model. 

\subsection{Metallicity Trends with Stream Coordinates}
\label{subsec:trends}

\begin{figure*}
    \includegraphics[width=\textwidth]{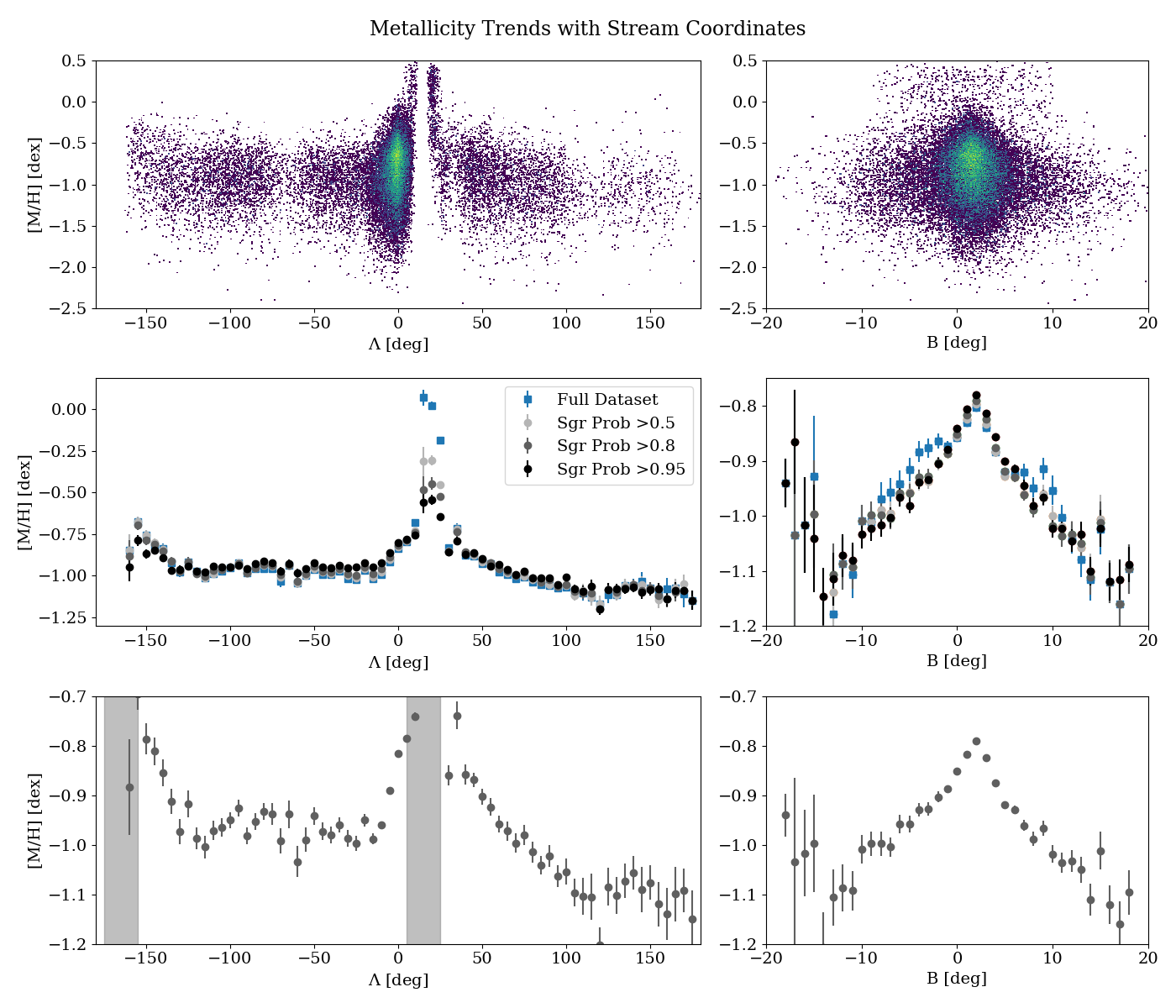}
    \caption{Metallicity trends as a function of stream longitude $\Lambda$ (left panels) and latitude $B$ (right panels). Top panels show the metallicities versus stream coordinates in a 2D histogram, while the middle and lower panels show the means in binned stream coordinates. Middle panel: the blue points in each figure show the full V21 sample binned in $\Lambda$ (left) and $B$ (right), with errorbars indicating the error on the mean in each bin. Grey and black points shown the resulting trends with different membership probability cuts imposed. The bins in $\Lambda$ nearest to the Galactic center are the most sensitive to the membership probability cuts. Bottom panels: the trends with stream coordinates for all stars with Sgr membership probability greater than 0.8. Grey bars indicate the approximate locations of the Galactic center ($\Lambda \sim 15 \degree$) and anti-center ($\Lambda \sim -165 \degree$). We note we have excluded the higher metallicity stars in the bottom left panel, which we expect to be primarily disk stars, to highlight the behavior of the mean metallicity along the stream. The gradient with stream latitude $B$ is not strongly affected the by membership probability cuts.}
    \label{fig:1dres}
\end{figure*}

\begin{figure*}
\includegraphics[width=\textwidth]{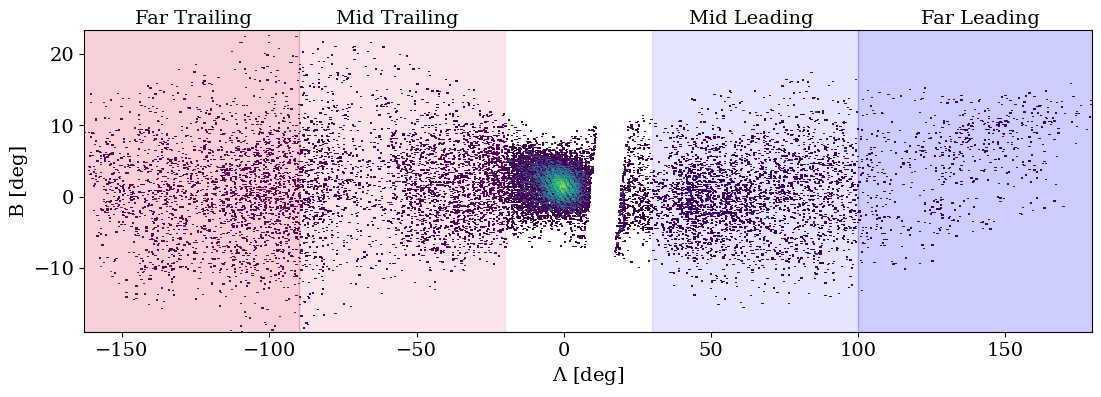}
\includegraphics[width=\textwidth]{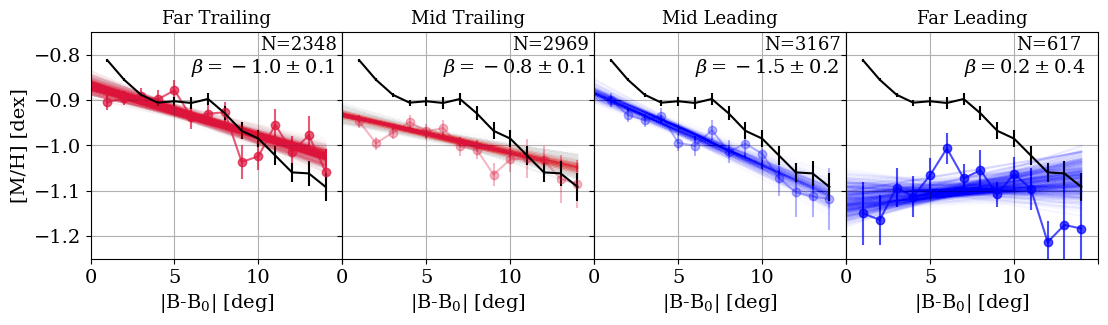}
    \caption{Gradients in stream latitude $B$ when the stream is divided by longitude $\Lambda$. In the top panel, we divide the stream into four components: far trailing ($\Lambda<-90\degree$), mid-trailing ($-90 \degree < \Lambda < -20 \degree$), mid-leading ($30 \degree < \Lambda < 100 \degree$) and far leading ($\Lambda>100 \degree$). The trends with $|B-B_0|$ for each stream component binned in $B$ are plotted in the lower panel. We fit for the slope of the gradient using the same mixture model described in Section \ref{sec:methods}, though model instead the slope as a function of $|B-B_0|$. Draws from the posterior for the gradient slopes are shown by the colored lines.  Gradient slopes are printed in each panel, in units of $10^{-2}$ dex/deg. While the gradients are not as steep as the gradient for the full sample (black lines in lower panels), there are still gradients in each stream component, except for the far leading component.}
    \label{fig:b_grad_lambda}    
\end{figure*}

Figure \ref{fig:1dres} shows the trends in [M/H] as a function of stream longitude $\Lambda$ (left panels) and stream latitude $B$ (right panels). The top panels represent the data as a 2D histogram, and the lower panels show the means of the data binned in stream coordinates. In the left panels, we can again see contamination from the metal-rich disk, as the high metallicity sample near the galactic midplane (identifiable by the gap in the top panels, reflecting the Galactic latitude cut imposed by V21). The gray bars in the bottom panels indicate the approximate location of the Galactic center ($\Lambda \sim 15 \degree$) and the Galactic anti-center ($\Lambda \sim 165 \degree$); we can see the mean metallicity as a function of $\Lambda$ increase near both these locations.

The middle panels of Figure \ref{fig:1dres} highlight the sensitivity of these observed trends to contamination. The blue squares in the middle panels of Figure \ref{fig:1dres} show the binned data for the full V21--A23 cross-matched sample, while the grey points show the binned data with increasingly stringent cuts on Sgr membership probability. Without modeling the contamination from the MW disk, as a function of $\Lambda$, the mean metallicity is peaked at $[M/H] \sim 0.17$ at $\Lambda=18\degree$; the mean metallicities over $5<\Lambda<25$, where the stream intersects the MW disk, decrease as the membership probability cut becomes increasingly strict. As a function of stream latitude $B$, we see that the mean metallicities are more robust to contamination, and the estimates derived after enforcing different membership probability cuts are consistent with one another. The mean metallicity of the peak increases slightly as the membership probability cut becomes more stringent; this is because of the metal-poor stars (that have higher probability of belonging to the outlier distribution) are removed from the sample, increasing the overall mean metallicity. We note that some of these stars classified as having higher outlier probabilities are likely in the metal-poor tail of the Sgr dwarf galaxy. 

In the lower panels of Figure \ref{fig:1dres}, we show the binned trends with stream coordinates only for stars with Sgr membership probabilities greater than 0.8, and zoom in to focus on the behavior of the mean metallicity along the stream. As a function of stream longitude $\Lambda$, the mean metallicity at the beginning of the trailing arm is high, which we expect is due to stars in the Galactic anti-center. We expect the contamination from the outer disk to extend to higher Galactic latitudes than near the Galactic center due to perturbations (e.g., \citealt{Laporte2022}). We then see that the mean metallicity is roughly constant along $-140 \degree < \Lambda < 40 \degree$, before sharply increasing near the progenitor (at $\Lambda=0$). The mean metallicity continues to increase sharply near the Galactic midplane due to disk contamination (see middle panels). As we transition to the leading arm, we see a steeper decrease in metallicity over $25 \degree< \Lambda < 100\degree$, at which point the metallicity is approximately constant (and is the most metal-poor along the stream). 

From Figure \ref{fig:1dres}, it is apparent that fitting linear relationships between stream longitude and metallicity is not the best description of the data (in the next section, we demonstrate that this is true even if Sgr has a linear radial metallicity gradient prior to disruption). However, to put our work in context with previous studies estimating a stream gradient, we fit a line to the mean metallicity as a function of stream longitude in the trailing and leading arms (see Equation \ref{eqn:lambda_only}). In the trailing arm ($\Lambda<0$), we find the slope to be $\alpha_T= (-1.25 \pm 0.05) \times 10^{-3}$ dex/deg. In the leading arm ($\Lambda>0$), we find the slope to be $\alpha_L=( -1.89 \pm 0.07) \times 10^{-3}$ dex/deg, steeper than the slope in the trailing arm. These gradients are shallower than those inferred by \cite{Hayes2020}, who find a slope in [Fe/H] of $-2.6 \times 10^{-3}$ dex/deg in the trailing arm and $-4.0 \times 10^{-3}$ dex/deg in the leading arm (when they include Sgr core stars in their fits, as we have also done here). While there are several possible reasons for this, especially given that the stream coverage in $\Lambda$ is very different across the two samples, one contributing factor could be the fact that the XGBoost appears to overestimate the metallicities for metal-poor stars and underestimates the metallicities for metal-rich stars (see Figure \ref{fig:apogee_v_XGBoost}). To test how much the over/under-estimation of the XGBoost metallicities is responsible for this discrepancy, we adjust the XGBoost metallicities by a ''correction factor" based on the line shown in Figure \ref{fig:apogee_v_XGBoost}. We find that the slopes steepen to $-1.7 \times 10^{-3}$ dex/deg in the trailing arm and $-2.5 \times 10^{-3}$ dex/deg in the leading arm, still significantly shallower than the gradients inferred by \cite{Hayes2020}. We therefore primarily attribute this difference to the different sample sizes and the different stream coverage between the studies. 

As a function of stream latitude $B$, we see a clear gradient in metallicity, peaked at the location of the main body ($B_0=1.5 \degree$). This gradient has not been detected before in other samples.  
When we model the mean metallicity only as a function of stream longitude $B$ (Equation 4), we find that the slope above the stream track ($B>B_0$) is $\beta_N= -0.0248 \pm 0.008$ dex/deg and below the stream track ($B<B_0$) is $\beta_S= -0.0202 \pm 0.008$ dex/deg. We note that these slopes are a factor of 10 steeper than the slopes found as a function of stream longitude. 

The final column in Table \ref{tab:results} summarizes the resulting gradient slopes when we allow for the mean metallicity to vary as a function of both stream coordinates (Equation \ref{eqn:lambda_beta}). These slopes are all shallower compared to the slopes when we model the gradient as a function of only one stream coordinate (left and middle columns of Table \ref{tab:results}). This is a result of the fact that the mean metallicity along the stream is a function of both coordinates (as opposed to just one), but all slopes are inconsistent with zero gradient.  

\begin{table*}[]
    \centering
    \begin{tabular}{c||ccc}
    & $\mu_{\rm Sgr} =\mu_{\rm Sgr} (\Lambda)$ & $\mu_{\rm Sgr} =\mu_{\rm Sgr} (B)$  &
  $\mu_{\rm Sgr} =\mu_{\rm Sgr} (\Lambda, B)$  \\
  \hline \hline
$\alpha_T$ [$10^{-3}$ dex/deg] & $-1.25 \pm 0.05$  &  -- & $-0.63 \pm 0.06$ \\
$\alpha_L$ [$10^{-3}$ dex/deg] & $-1.89 \pm 0.07$ & -- & $-1.25 \pm 0.08$ \\
$\beta_S$  [$10^{-2}$ dex/deg] &  -- & $-2.02 \pm 0.08$ & $-1.33 \pm 0.09$ \\
$\beta_N$ [$10^{-2}$ dex/deg] & -- & $-2.48 \pm 0.08$ & $-1.78_{-0.10}^{+0.09}$ \\
$C$ [dex] & $-0.831 \pm 0.002$& $-0.815 \pm 0.002$ & $-0.810 \pm 0.003$ \\
$\sigma_{\rm Sgr}$ [dex] & $0.289 \pm 0.002$ & $0.286 \pm 0.002$ &$0.284 \pm 0.002$\\
$f_{\rm Disk}$ [\%] & $2.2 \pm 0.1$ &$2.4 \pm 0.1$ & $2.4 \pm 0.1$\\
$f_{\rm Outlier}$ [\%] & $2.1 \pm 0.2 $ &$2.3 \pm 0.2 $ & $2.4 \pm 0.2 $ \\
    \end{tabular}
    \caption{Resulting parameter estimates from our inference procedure for our three parameterizations of the metallicity variation in the Sgr stream. We quote the medians of our posterior distributions, and uncertainties are the 16/84 percentiles. The first column lists the parameter estimates when modeling the mean metallicity as a function of only longitude ($\mu_{\rm Sgr} =\mu_{\rm Sgr} (\Lambda)$); the second column contains results for the gradients only as a function of stream latitude ($\mu_{\rm Sgr} =\mu_{\rm Sgr} (B)$); and the resulting gradients when the metallicity varies as a function of both stream coordinates ($\mu_{\rm Sgr} =\mu_{\rm Sgr} (\Lambda, B)$ ) are in the third column.}
    \label{tab:results}
\end{table*}

We further explore the longitude ($\Lambda$) dependence of the latitude ($B$) metallicity gradient in Figure \ref{fig:b_grad_lambda}. We divide the stream into four quadrants, excluding the region immediately surrounding the main body and the MW disk. In the lower panels of Figure \ref{fig:b_grad_lambda}, we compare the trend of [M/H] with $|B-B_0|$ (where $B_0=1.5 \degree$ is the location of the core). The black line in each panel is the gradient measured over the full sample, while the colored points each represent a different division of the stream. We denote the ``Far Trailing" component to have $\Lambda<-90\degree$; the ``Mid Trailing" component has $-90 \degree < \Lambda < -20 \degree$; the ``Mid Leading" component has $30 \degree < \Lambda < 100 \degree$; and the ``Far Leading" component has $\Lambda>100 \degree$. For the purposes of this analysis, we exclude a relatively large, conservative region near the core and the galactic center to ensure that the trends are dominated by stream stars.

Within each stream component, we model the slope of the gradient with respect to $|B-B_0|$, where we transition to focusing on the absolute value of stream latitude in order to get the most signal from our now smaller samples. Colored lines show draws from the posterior estimates for the gradients. Posterior distribution medians for the slope of the gradients $\beta$ are quoted in each panel of Figure \ref{fig:b_grad_lambda}, in units of $10^{-2}$ dex/deg. 

As is to be expected, the gradients are not as steep within the stream components compared the measured gradient using the full sample. Given that the most metal-rich stars are in the core (e.g., Figure \ref{fig:map}), the peak metallicity in the different stream components is expected to be lower. However, we find significant gradients with $|B|$ (i.e., inconsistent with zero by $7-8 \sigma$) in the far trailing, mid trailing and mid leading components. Interestingly, the steepest gradient seen in the mid-leading portion of the stream, which is also the region of the stream where the metallicity drops off most steeply with $\Lambda$. 

In the far leading component of the stream, we find a gradient slope consistent with zero. This could be partially due to the fact that there are fewer stars in this component of the stream compared to the other quadrants by a factor of a few, so our estimates of the gradient slope will be more uncertain compared to the other quadrants. 
In addition, several N-body models of the stream find that the far portion of the leading arm is primarily composed of stars that are stripped earlier compared to the far trailing arm (i.e., the median stripping time decreases more quickly with stream longitude in the leading arm than in the trailing arm; e.g., \citealt{LM2010}, V21; see also Figure 10 in \citealt{Hayes2020}). If the debris in this quadrant was primarily stripped earlier compared to the other quadrants, we would expect it to contain the fewest metal-rich stars and have the shallowest metallicity gradient.  

\begin{figure*}
\includegraphics[width=\textwidth]{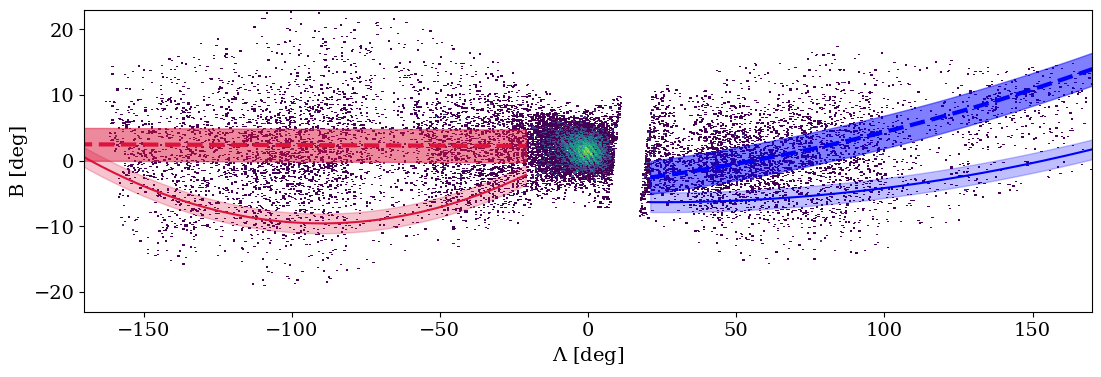}
\includegraphics[width=\textwidth]{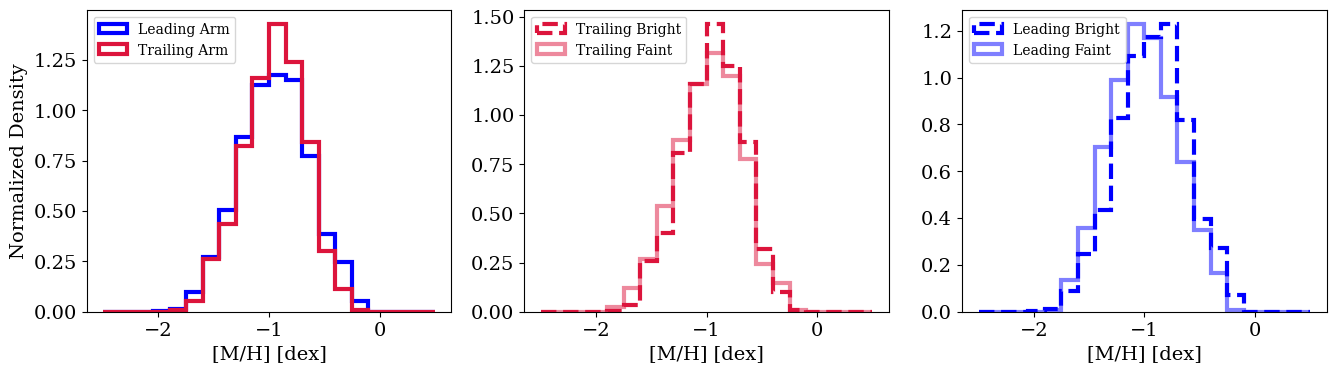}
    \caption{Top panel: the Sgr sample overlaid with the tracks of the Sgr bifurcation as defined in \cite{Ramos2022}. Lower left: comparisons of the MDFs of the leading (blue) and trailing (red) arms; they are consistent with one another. The lower middle and left panels compare the MDFs for the bright and faint branches for the trailing and leading arms, respectively. While the two branches are essentially identical in the trailing arm, in the faint branch is slightly more metal-poor than the bright branch in the leading arm.}
    \label{fig:bif}    
\end{figure*}

\subsection{Bifurcation}
\label{subsec:bifurcation}

One significant puzzle plaguing studies of Sgr is the stream bifurcation: the stream appears to have both a primary bright track and a fainter track that is misaligned on the sky with the bright track. This bifurcation has been observed in both the leading and trailing arms (\citealt{Belokurov2006}, \citealt{Koposov2012}, \citealt{Ramos2022}). 

One hypothesis for the origin of the bifurcation is that the Sgr progenitor had a disk. \cite{Penarrubia2010} showed that a model of a rotating disk that is misaligned with the orbital plane of the satellite could produce the stream bifurcation; however, the rotation of the progenitor implied by this configuration was not consistent with later observations of the Sgr remnant \citep{Penarrubia2011}. Recently, \cite{Oria2022} revisited this scenario with a series of simulations with the same initial conditions as V21 but with test particles on disk-like orbits on different inclinations, and found that a spiral disky structure (oriented nearly perpendicular to the Sgr orbital plane and the MW disk plane) can reproduce the properties of the faint branch without requiring significant rotation in the core.

We compare the MDFs of the leading and trailing arms, as well as the bright and faint branches of the bifurcation, in Figure \ref{fig:bif}. The top panels show the tracks for the bright and faint branches as traced in \cite{Ramos2022} overlaid on the stream density. We note that we are using a different coordinate convention from \citep{Ramos2022}; the sign of $B$ is flipped here compared to their convention. The lower panels of Figure \ref{fig:bif} show the MDFs for each population. For the purposes of these figures, we exclude all stars with $|\Lambda|<20$ deg so that the samples are not dominated by stars in the main body, and only include stars with Sgr membership probability greater than 0.8. This leaves us with 5179 stars in the trailing arm and 4056 in the leading arm. 

The lower left panel in Figure \ref{fig:bif} compares the MDFs of the leading (purple) and trailing (orange) arms; the MDFs for the two populations are very similar, and the means of both samples are $-0.95$ dex. In other studies of the Sgr stream with spectroscopic abundances, the trailing arm is usually found to be more metal-rich on average than the leading arm (see, e.g., \citealt{Yang2019,Hayes2020,Ramos2022, Limberg2023}). However, in these works, the samples in the leading arm usually start at larger $\Lambda$, and do not have as many stream stars closer to the Sgr main body as the trailing samples. Figure \ref{fig:bif} demonstrates that including stars closer to the main body in Sgr in the leading arm produces samples with consistent overall MDFs; however, as seen in the previous section, the trends with stream longitude are different in the two arms, with [M/H] decreasing more steeply with $\Lambda$ in the leading arm than the trailing arm.

In the middle and right panels of Figure \ref{fig:bif}, we weight each point in the histogram according to its probability of belonging to the faint or bright branches, as defined based on the tracks measured in \cite{Ramos2022}. \cite{Ramos2022} describe each track as a Gaussian of constant width, with $\sigma=1.5\degree$ for the faint track and $\sigma=2.5\degree$ for the bright track (though they note this is a simplifying assumption). These tracks are traced out in the top panel of Figure \ref{fig:bif}. For the trailing arm, the average metallicity of the faint branch is $0.03$ dex more metal poor than the bright branch. In the leading arm, the faint branch is slightly more metal poor than the bright branch, with the means offset by $0.07$ dex (consistent with the findings from \citealt{Ramos2022}), but this difference is also small given the measurement uncertainties of $\sim 0.1$ dex. 

However, we note that the faint branches of the bifurcation are tracing out debris that is, on average, at higher latitudes from the stream track. Therefore, the differences in the metallicities in the different branches could be a reflection of the fact that there is a metallicity gradient across the stream. As we will show in the next section, an N-body model of the stream does not need to produce the bifurcation in order to have a metallicity gradient with stream latitude. Future simulations similar to those performed in \citep{Oria2022} with painted metallicities could provide further evidence in favor of or against the scenario in which the bifurcation is due to an initially disky Sgr dwarf galaxy progenitor. In the next section, we investigate the predicted trends for metallicity with stream coordinates in the case where Sgr is initialized as a spheroid. 

\section{Comparison with Simulations}
\label{sec:sims}

In this work, we report a metallicity gradient in the Sgr stream perpendicular to the stream track for the first time. In order to determine if this is expected from theoretical predictions, we compare our results with N-body simulations of the disruption of a Sgr-like system. We introduce the simulations in Section \ref{subsec:sims} and report trends with initial internal binding energy and initial radius with stream coordinates in Section \ref{subsec:sim_trends}. To compare with observational studies of local dwarf galaxies, in Section \ref{subsec:paint} we paint metallicity gradients on to the simulated dwarf to show what metallicity trends with stream coordinates would be expected with different initial metallicity gradients.

\subsection{Simulations}
\label{subsec:sims}

\begin{figure*}
\centering
    \includegraphics[width=\textwidth]{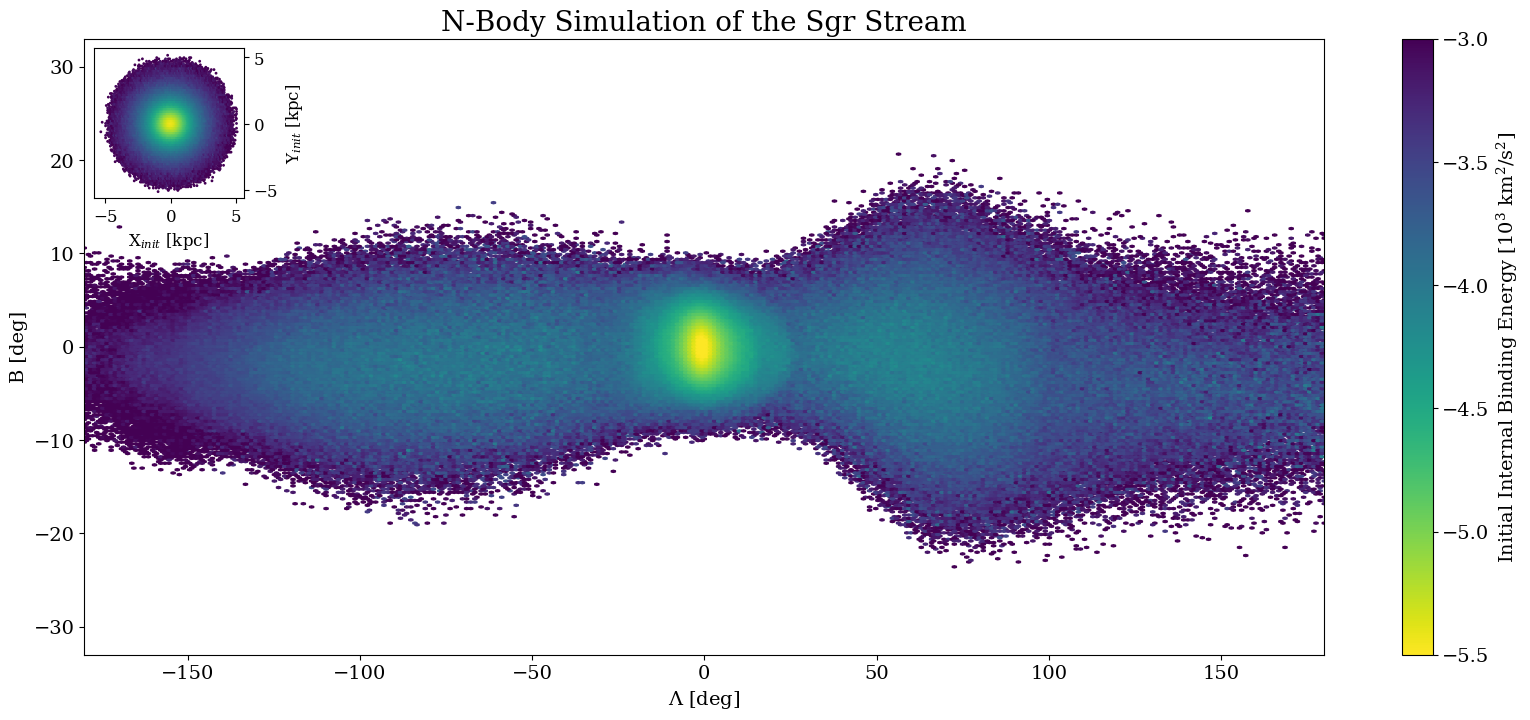}
    \caption{Simulation of the Sgr stream, color-coded by mean initial internal binding energy within the satellite prior to disruption ($E_{\rm Bind}$). The particles that were the most tightly bound in the Sgr progenitor dwarf galaxy are located in the center (where we observe the main body of the Sgr dwarf spheroidal today). The mean $|E_{\rm Bind}|$ decreases with stream longitude $\Lambda$, but also with stream latitude $B$. Inset: initial $X-Y$ positions of particles colored by their initial binding energy, shown on the same scale as the main figure. Initial internal binding energy scales with initial radius within the progenitor.}
    \label{fig:sim_map}
\end{figure*}

\begin{figure*}
\centering
    \includegraphics[width=0.9\textwidth]{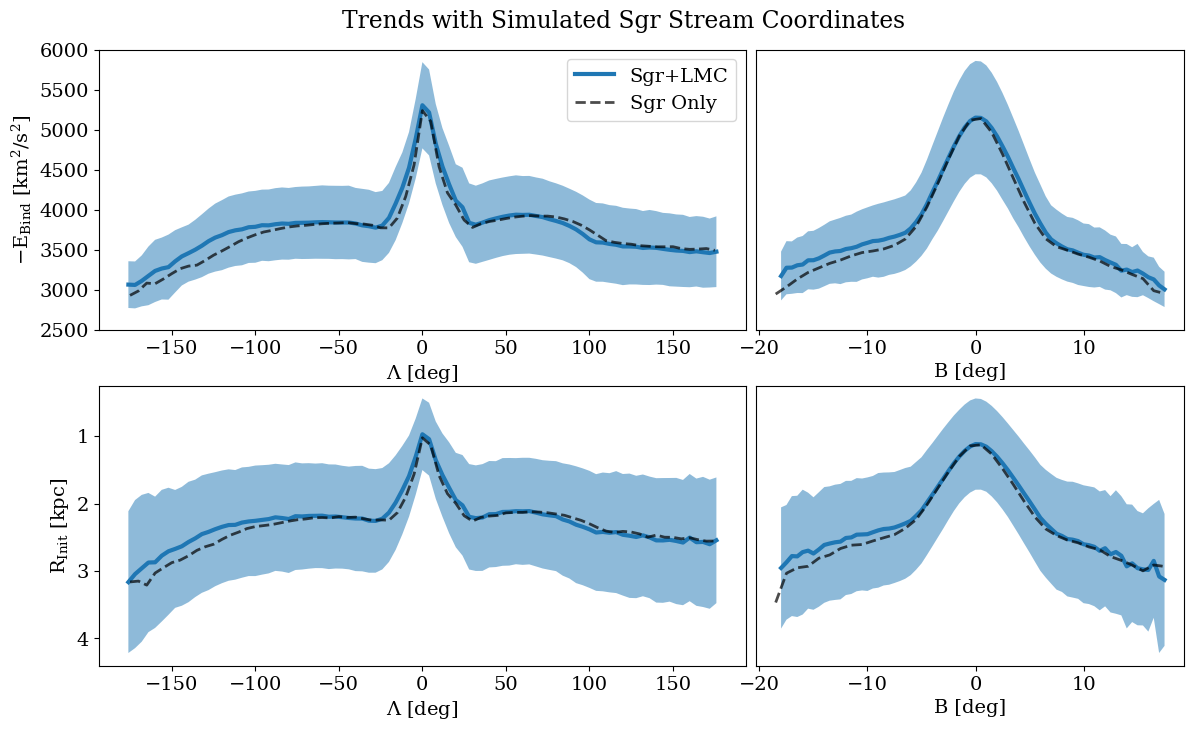}
    \caption{Trends in mean initial internal binding energy ($E_{\rm Bind}$; top panels) and initial radius within the progenitor (bottom panels) binned in stream longitude $\Lambda$ (left panel) and stream latitude $B$ (right panel). Blue regions indicate the 1$\sigma$ spread about the mean. We note that there is not a linear relationship between $-E_{\rm Bind}$ and initial radius with either stream coordinate, and that the trends shown here are remarkably similar to those seen in the data in Figure \ref{fig:1dres}. The blue curves show the trends for the \cite{Stelea23} simulation including the LMC, while the gray dashed line shows the trends for the simulation only including Sgr; the differences between the two are subtle.}
    \label{fig:sim_means}
\end{figure*}

\begin{figure*}
    \centering
    \includegraphics[width=\textwidth]{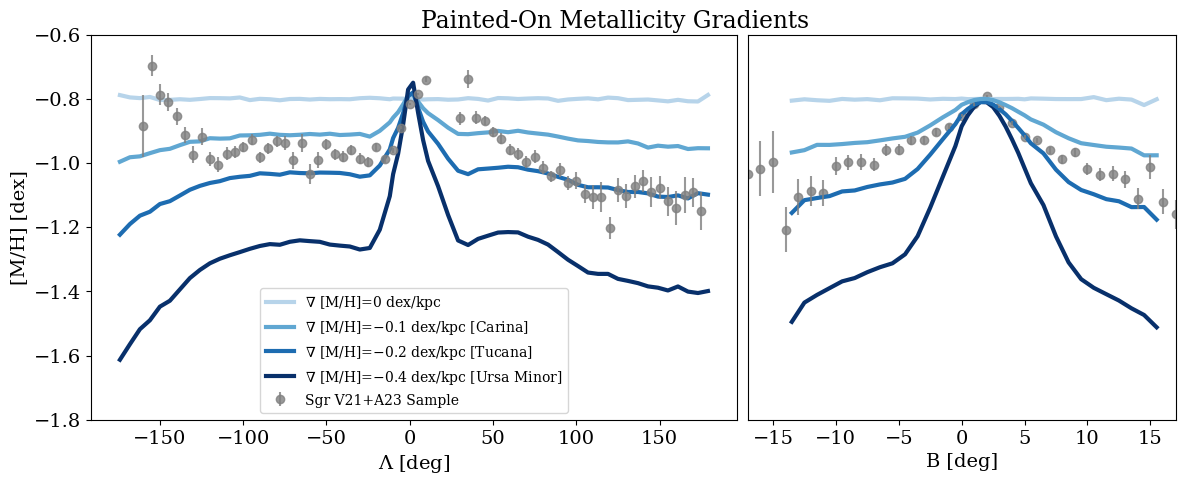}
     \caption{Metallicity trends in longitude (left) and latitude (right), for the \cite{Stelea23} simulations with different metallicity gradients. Colored lines show four different choices of initial metallicity gradients consistent with those observed in local dwarf galaxies, where we have matched the peak metallicity as a function of stream latitude to the data. The observed trends with metallicity are consistent with an initial gradient $\sim -0.1$ to $-0.2$ dex/kpc, well within the observed range of local dwarfs.}
    \label{fig:sim_gradients}
\end{figure*}

In this section we briefly describe the setup and evolution of the MW-like disk galaxies that we use as the host galaxies for the merger simulations, the satellite used to model Sgr, and their combination. These models are presented primarily in \cite{Stelea23}, where the setup is described in full. We make use of their MW + Sgr model, and their MW + Sgr + LMC model.

The initial conditions for the MW-like host galaxy, the LMC and Sgr are all generated using the galactic dynamics library \texttt{Agama} \citep{Vasiliev-agama}. The host galaxy model broadly follows the axisymmetric halo model of \cite{Vasiliev2021}, although \cite{Stelea23} use a distribution function based version of the initial condition generator, and reduce the disk velocity dispersion. The halo, disk and bulge components have a total mass of $7.33\times10^{11}\ M_{\odot}$, $5\times10^{10}\ M_{\odot}$ and $1.2\times10^{10}\ M_{\odot}$ respectively. This is a light MW halo compared to most estimations in the literature, but was shown to provide a good fit to the Sgr stream in \cite{Vasiliev2021} which is sufficient for the illustration in this work.

The dwarf galaxy used for Sgr is taken directly from \cite{Vasiliev2021}. It consists of two spherical components to model the stellar and dark matter. The stellar component consists of $4\times10^6$ particles and follows a King profile \citep{King62} with scale radius 1 kpc, King parameter $W_0$=4, and a total initial mass of $2\times10^{8}\ M_{\odot}$. The dark matter is a spheroid consisting of $6.4\times10^{6}$ particles with scale radius 8 kpc, and a total initial mass of $3.6\times10^{9}\ M_{\odot}$. The LMC is modeled simply as a single spheroid consisting of $2\times10^6$ particles, with scale radius 10.8 kpc and a total mass of $1.5\times10^{11}\ M_{\odot}$. 

The models are evolved with the GPU based $N$-body tree code \texttt{Bonsai} \citep{Bonsai,Bonsai-242bil} for $\sim3$ Gyr, using a smoothing length of 20 pc and an opening angle $\theta_{\mathrm{o}}=0.4$ radians. Table \ref{sgrpos} shows the initial Galactocentric position and motion of the model Sgr along with the present day position and motion of the remnant.

\begin{table*}
\caption{Sgr remnant position and motion from the literature, and the initial and `Present day' position of the Sgr model from \citet{Stelea23}.}
\begin{tabular}{@{}lllllllll@{}}
\toprule
Model/Catalogue & $x$ (kpc) & $y$ (kpc) & $z$ (kpc) & $v_x$ (km s$^{-1}$) & $v_y$ (km s$^{-1}$) & $v_z$ (km s$^{-1}$) \\ 
\midrule
Sgr \citep{Vasiliev2020} & 17.5 & 2.5 & -6.5 & 237.9 & -24.3 & 209.0 \\
\midrule
Sgr initial & 66.3 & 2.4 & -26.2 & 9.8 & -46.0 & 91.9 \\
Sgr final (With LMC) & 16.2 & 2.3 & -6.5 & 240.8 & -89.5 & 183.6 \\
Sgr final (Without LMC) & 18.5 & 2.6 & -12.3 & 206.5 & -108.5 & 151.5 \\ 
\bottomrule
\end{tabular}
\label{sgrpos}
\end{table*}

We emphasize that these simulations are not tuned to perfectly reproduce the properties of the stream, and we are therefore cautious in making quantitative comparisons with these simulations to our data. However, these simulations are ideal for demonstrating for how we might expect the stellar metallicity of Sgr to vary given different initial metallicity gradients. 

\subsection{Trends with Initial Binding Energy}
\label{subsec:sim_trends}

To build intuition for how we might expect metallicity to vary as a function of stream coordinates, we highlight how stars are distributed along and across the stream as a function of their initial internal binding energy ($E_{\rm Bind}$) within the satellite prior to disruption. The internal binding energy is the sum of the gravitational potential and kinetic energy of the particle within the satellite before disruption by the MW: $E_{\rm Bind}=\Phi_{\rm Sat}+\frac{1}{2}v_{\rm Sat}^2$. While metallicity gradients are observed in dwarf galaxies as a function of radius, in the simulation space, it can be useful to think instead of metallicity varying as a function of internal binding energy.  A star's radius from the center of the progenitor will change as a function of its orbital phase (and will change dramatically if the star is on an eccentric orbit), while its internal binding energy is conserved (until the progenitor begins to interact with the MW-like host). Systems with observed metallicity gradients with radius will therefore also have gradients with respect to initial internal binding energy. In the below discussion and figures we refer primarily to $-E_{\rm Bind}$, as this is the quantity we expect to be positively correlated with metallicity (as, by definition, $E_{\rm Bind}<0$ for bound particles). 

Figure \ref{fig:sim_map} shows the Sgr stream from the \cite{Stelea23} simulations (here showing the model including the LMC) color-coded by initial internal binding energy within the satellite (with the inset highlighting how $E_{\rm Bind}$ varies as a function of initial position within the progenitor). We note that in making these comparisons, we have fully unwrapped the stream and only focus on the debris within $\pm 180 \degree$ from the progenitor, deliberately ensuring that overlapping debris from previous wraps are not the source of any observed gradients here\footnote{While we do not show them here, we note that we also computed mean $-E_{\rm Bind}$ without unwrapping the stream, and found that the differences in the trends with stream coordinates and $-E_{\rm Bind}$ were very small.}. Figure \ref{fig:sim_means} shows the trends with $-E_{\rm Bind}$ as a function of stream longitude (left panel) and latitude (right panel), and trends with initial radius are shown in the lower two panels. The shaded regions encompass the $1\sigma$ spread in each bin of stream longitude/latitude. We also include the model with no LMC for comparison, shown by the grey dashed lines; the differences between the two are very subtle. While the exact locations and velocities of the remnants are slightly different across the two simulations, we note that we have centered the coordinate system on the remnant for the purposes of these figures. We immediately see similarities in these trends with how metallicity is varying along and across the Sgr stream in observations as seen in Figure \ref{fig:1dres}. As a function of stream longitude $\Lambda$, stars with larger $|\Lambda|$ values were less tightly bound to the Sgr progenitor. The mean initial $-E_{\rm Bind}$ increases smoothly from $\sim -180 < \Lambda <  -100$ and then is roughly constant until approaching the core, at which point it sharply increases and sharply decreases around the core. 
We see a similar trend in stream latitude $B$, with $-E_{\rm Bind}$ dropping off more quickly as a function of $|B|$ than as a function of $|\Lambda|$. 

How would we expect these trends to translate to observed metallicities? In order to directly compare with our observations, in the next section, we ``paint" metallicities on to the simulation.

\subsection{Painting on Gradients}
\label{subsec:paint}

The \cite{Stelea23} simulations are a suite of N-body only simulations; therefore, there is no star formation or chemical enrichment. To place our observed metallicity gradients in Sgr in context with observed metallicity gradients in local dwarf galaxies (e.g., \citealt{Taibi2022} and references therein), we ``paint" metallicities onto the star particles from the \cite{Stelea23} simulation. We consider four choices of initial radial gradients, each with a scatter of 0.3 dex, normalized such that the mean metallicity at the progenitor matches the observed metallicity of the core (we match the peak in $B$ rather than $\Lambda$, as the peak $\Lambda$ is more sensitive to disk contamination; see Figure \ref{fig:1dres}). We consider a completely flat gradient as well as gradients of $-0.1$ dex/kpc (consistent with, e.g., Carina; \citealt{Koch2006}), $-0.2$ dex/kpc (similar to Tucana; \citealt{Taibi2020}), and $-0.4$ dex/kpc (as observed in Ursa Minor; \citealt{Pace2020}). Figure \ref{fig:sim_gradients} shows the results of this analysis: the mean metallicity as a function of stream longitude (left) and latitude (right) are shown for the data (grey) and increasingly steep gradients (in increasingly dark shades of blue). 
We can see immediately from Figure \ref{fig:sim_gradients} that a linear relationship between [M/H] and initial radius does not result in a linear relationship between [M/H] and $\Lambda-\Lambda_0$ or $B-B_0$ (where here we have centered $\Lambda_0, B_0 = (0 \degree, 0 \degree)$).  

A flat metallicity gradient in the progenitor does not reproduce any of the trends with stream coordinates that we have observed here. Qualitatively, the observed gradient is consistent with an initial metallicity gradient between $-0.1$ to $–0.2$ dex/kpc, well within the range of observed gradients in Local Group dwarfs. This gradient is also much steeper than the recently inferred gradient for the disrupted system Gaia-Enceladus-Sausage ($\sim -0.02$ dex/kpc; \citealt{Chandra2022}). 

As this model is not tuned to fit the properties of the stream, we refrain from making more specific quantitative claims about the metallicity gradient of the Sgr progenitor in this work. Furthermore, if the progenitor of Sgr had a disk (e.g., \citealt{Penarrubia2010}, \citealt{Oria2022}), we would expect the metallicity trends as a function of stream coordinates to depend on the inclination of the disk with respect to the orbital plane. 
Our results are therefore a novel constraint for models of the internal structure of the Sgr progenitor to be tested in future theoretical studies.

\section{Conclusions}
\label{sec:concl}

In this paper, we map the chemical structure of the Sgr stream using $\sim$34,000 stars with metallicity estimates derived from Gaia XP spectra. Our main results are as follows:

\begin{enumerate}
    \item We present a catalog of 34,240 RGB stars that are high probability Sgr stream members (identified in V21) and their chemical abundances inferred from Gaia XP Spectra (in A23) (see Figure \ref{fig:map}). This is the largest sample of Sgr stream stars to date with chemical abundance information. While at the fainter end of the Gaia sample with available XP spectra, the inferred abundances from XGBoost are generally consistent with Sgr stars that also have APOGEE spectra, though we note that the XGBoost models appear to slightly over (under) estimate the metallicities of low (high) metallicity stars. This could imply that the slopes of the gradients reported here are lower limits on the true gradient slopes.
    \item We find a clear metallicity gradient as a function of stream latitude $B$. Above the stream track ($B>B_0$), we find $\nabla \mathrm{[M/H]} = -2.48 \pm 0.08 \times 10^{-2}$ dex/deg, $\nabla \mathrm{[M/H]} = -2.02 \pm 0.08 \times 10^{-2}$ dex/deg below the stream track ($B<B_0$). This is the first detection of a metallicity gradient perpendicular to the stream track. 
    \item Dividing the stream into different longitude bins, we find that the stream has a metallicity gradient with respect to stream latitude $B$ in all segments, except for in the far leading arm ($\Lambda>100 \degree$). The slopes of the gradient are not as steep as the measured slope when the Sgr main body is included, which is to be expected as the main body contains the highest metallicity stars. However, the slopes in the far trailing, mid-trailing, and mid-leading arms are all inconsistent with zero. 
    \item We also measure metallicity gradients in the leading and trailing arms, though we note that a linear gradient as a function of stream longitude does not appear to be a good model for the data. Instead, we find the metallicity to remain roughly constant along the trailing arm, with a steep increase near the core at $\Lambda=0$. In the leading arm, the mean metallicity falls off more steeply than in the trailing arm, before leveling off to a relatively constant lower metallicity. This asymmetry is to be expected based on how distance varies along the stream. 
    \item We compare our findings with N-body models for the Sgr stream. We show how the initial internal binding energy (and initial radii) of stars within the satellite varies with stream longitude and latitude, and find trends consistent with our data. We highlight how a linear metallicity gradient within the Sgr dwarf galaxy progenitor does not result in a linear relationship between stream coordinates and metallicity. 
    \item Finally, we paint different initial radial metallicity gradients onto our simulation and compare with the data. 
    We find that the observed metallicities are consistent with an initial metallicity gradient in the Sgr progenitor of $\sim -0.1$ to $-0.2$ dex/kpc, consistent with metallicity gradients observed in local dwarf galaxies. While the simulation used here initializes Sgr as a spheroid, future work comparing to simulations where Sgr is initialized as a disk is needed to determine if our gradients are consistent with a disky Sgr progenitor. 
\end{enumerate}

While the Gaia XP inferred abundances are very uncertain for individual stars, our results highlight the information available from leveraging the vast number of stars with available XP spectra. Our results are promising for using Gaia XP inferred stellar parameters (and future abundances from low-resolution spectra) for characterizing the properties of the disrupted dwarf galaxies that comprise the stellar halo. Future higher resolution spectroscopic surveys providing information about more elements, along with future simulations including gas (e.g., \citealt{Tepper2018}) as well as self-consistent star formation and chemical enrichment, will further tighten constraints on the properties of the Sgr dwarf galaxy progenitor. Our results show that the observed gradients will have significant constraining power for the properties of the progenitor when compared to future tailored simulations. 

\acknowledgments

ECC acknowledges support for this work provided by NASA through the NASA Hubble Fellowship Program grant HST-HF2-51502 awarded by the Space Telescope Science Institute, which is operated by the Association of Universities for Research in Astronomy, Inc., for NASA, under contract NAS5-26555. ECC would also like to thank Zachary Jennings, Alis Deason, Denis Erkal, the CCA Dynamics Group and the Columbia Milky Way Stars Group for helpful discussions. This work has made use of data from the European Space Agency (ESA) mission
{\it Gaia} (\url{https://www.cosmos.esa.int/gaia}), processed by the {\it Gaia}
Data Processing and Analysis Consortium (DPAC,
\url{https://www.cosmos.esa.int/web/gaia/dpac/consortium}). Funding for the DPAC
has been provided by national institutions, in particular the institutions
participating in the {\it Gaia} Multilateral Agreement.
KVJ thanks the Center for Computational Astrophysics (Flatiron Institute) and the Simons Foundation for their past and ongoing support. IE acknowledges generous support from a Carnegie-Princeton Fellowship through Princeton University and the Observatories of the Carnegie Institution for Science.

\software{Agama \citep{Vasiliev-agama}, Astropy \citep{astropy:2013, astropy:2018, astropy:2022}, Astroquery \citep{astroquery2019}, Bonsai \citep{Bonsai},  emcee \citep{Foreman-Mackey2013}, IPython \citep{Perez2007}, Matplotlib \citep{Hunter2007}, Numpy (\citealt{harris2020array}), Pandas \citep{mckinney-proc-scipy-2010}, Scipy \citep{2020SciPy-NMeth}}

\bibliography{refs}

\end{document}